\documentclass[12pt]{article}

\usepackage{epsfig,times,citesort}

\renewcommand{\theequation}{\arabic{equation}}

\def  \bcen  {\begin{center}}
\def  \ecen  {\end{center}}
\def  \eg    {{\sl e.g.},}
\def  \beq   {\begin{equation}}
\def  \eeq   {\end{equation}}
\def  \beqa  {\begin{eqnarray}}
\def  \eeqa  {\end{eqnarray}}

\begin{document}

\begin{titlepage}

\renewcommand{\thefootnote}{\fnsymbol{footnote}}

\begin{flushright}
{\tt hep-ph/0407050} 
\end{flushright}

\vskip 1cm

\bcen
{\Large \bf \boldmath $B_d - \bar{B}_d$ mass
difference in Little Higgs model} \\

\vskip 1.5cm

{\sc \sffamily
S. Rai Choudhury$^1$\footnote{src@physics.du.ac.in}, 
Naveen Gaur$^1$\footnote{naveen@physics.du.ac.in}, 
Ashok Goyal$^1$\footnote{agoyal@iucaa.ernet.in}, 
Namit Mahajan$^2$\footnote{nmahajan@mri.ernet.in}
}
\vskip 1cm
$^1$
{\sl Department of Physics \& Astrophysics, \\
University of Delhi, Delhi - 110 007, India.} \\[2ex]
$^2$
{\sl Harish-Chandra Research Institute, \\
Chattnag Road, Jhunsi, Allahabad - 211019, India.}
\ecen

\vskip 2cm
\begin{abstract}
\noindent An alternate solution of hierarchy problem in the Standard
Model namely, the Little Higgs model, has been proposed lately. In
this work $B_d^0-\bar{B_d^0}$ mass difference in the framework of the
Little Higgs model is evaluated. The experimental limits on the mass
difference is shown to provide meaningful constraints on the parameter
space of the model.    
\end{abstract}

\end{titlepage}

%%%%%%%%%%%%%%%%%%%%%%%%%%%%%%%%%%%%%%%

\setcounter{footnote}{0}

\hrule
\vskip .2cm
{\sl 
\noindent Note : In our earlier version of the paper a mixup of two
symbols in our numerical computation lead to incorrect values of the
$B_d - \bar{B}_d$ mass difference. A paper that appeared subsequently
by Buras {\sl et.al.} \cite{Buras:2004kq}, presented the correct
calculation of $B_d - \bar{B}_d$ mass difference in Little Higgs
model. In this version we have corrected our numerical mistake. Our
results for $B_d - \bar{B}_d$ mass difference now agrees with the
results of Buras {\sl et.al.} \cite{Buras:2004kq}. Some typos in our
earlier version have also been corrected.}   
\vskip .2cm
\hrule

\vskip .7cm

\par Our understanding of the Standard Model (SM) is plagued by a
major issue, the {\sl ``hierarchy problem''}, arising out of the enormous
difference between the electroweak and the Plank scale. For quite
some time, Supersymmetry had provided an elegant framework for
solving this
problem although to-date there is no compelling experimental evidence
in its support. During the last two years, an alternative possibility
has been introduced in the literature where the Higgs mass remains
small by virtue of it being a Goldstone boson of a global 
symmetry which is broken at a scale above the electroweak scale. These
models are generically called the {\sl ``Little Higgs''} models and the
simplest of these, the {\sl ``Littlest Higgs''}(LH) model
\cite{Arkani-Hamed:2001nc}, has the least number of additional
particles involved. 

\par In the gauge sector, the LH model contains weakly coupled gauge
bosons with masses in the TeV scale in addition to the SM $W^\pm$ and
$Z$ \cite{Han:2003wu,Schmaltz:2002wx}. These mix amongst themselves
causing modification of SM gauge 
couplings of $W^\pm, Z$ with fermions and among themselves. In the
quark sector, a vector-like heavy top quark comes into play with mass
in TeV range, 
which has trilinear coupling with SM gauge bosons. Once again the
heavy top quark has mixing possibility with the SM top quark, resulting in
modification of coupling structure of quarks with $W^\pm$ and
$Z$. In addition, the model has charged Higgs bosons which introduce
scalar couplings with quark. Also, a heavier photon with mass in the
TeV range emerges, which couples both to leptons and quarks. 

\par The presence of these new particles as well as changes in the
SM interaction vertices, can cause changes in a variety of measurable
parameters. Some of them have already been calculated in the
literature
\cite{Han:2003wu,Han:2003gf,Chen:2003fm,Huo:2003vd,Choudhury:2004ce}.
These results provide good constraints on the parameters 
entering the LH model. Direct experimental confirmation of several
aspects of LH, \eg the masses of the heavy t-quark and the doubly
charged Higgs, would require sharper estimates of the parameters of the
theory. It is desirable therefore, to work out the consequences of the
LH-model for as many observable quantities as possible in order to
sharpen the constraints on the parameter space of such a model. In this
note, we report on a calculation of $B_0 - \bar{B}_0$ and $K_0 -
\bar{K}_0$ mixing in the context of LH model.

\par In SM, there is one basic box diagram responsible for generating
the effective Hamiltonian for the mixing of $B_0 - \bar{B}_0$ and $K_0
- \bar{K}_0$. In LH, there are many more box diagrams (as shown in
Figure 1) to be evaluated. The couplings and propagators required for
calculating these diagrams are listed in \cite{Han:2003wu}. 

\par
\begin{figure}[h]
\begin{center}
\epsfig{file=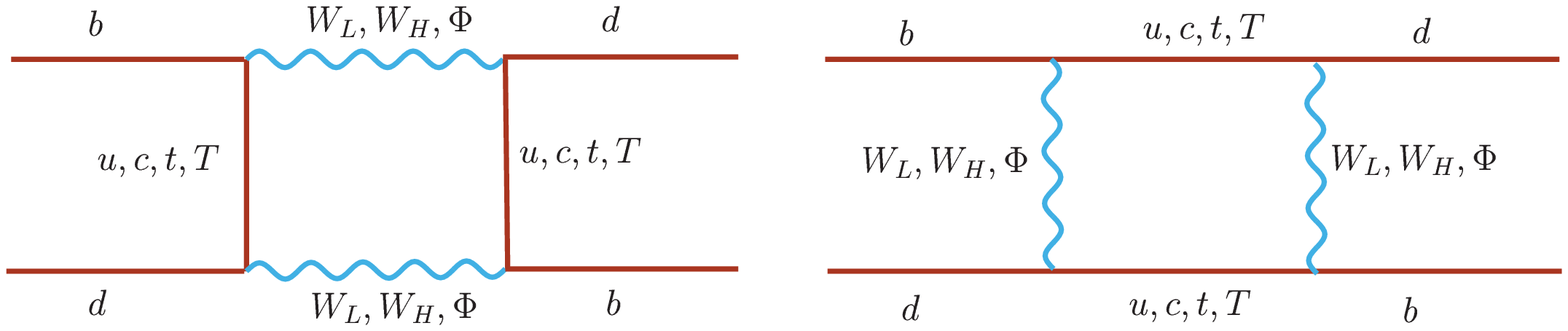,width=.9\textwidth,height=1.8in}
\caption{Box diagrams in LH. }
\label{fig:1}
\end{center}
\end{figure}

 The effective Hamiltonian resulting for the graphs in Fig.1 has
the structure :
\beq
{\cal H}_{eff} = \frac{G_F^2}{16 \pi^2} M_W^2 S_q (\bar{q}
d)_{V-A} (\bar{q} d)_{V-A}
\label{eq:1}
\eeq
with $q = b, s$ for ($B_0 - \bar{B}_0$) and ($K_0 - \bar{K}_0$) mixing
respectively. The invariant function $S_q$ has the following form:  
\beq
S_q = S_q^{SM} + S_q^{LH}
\label{eq:2}
\eeq
where in both $S_b$ and $S_s$, the first term represents the SM
contribution along with QCD corrections which are given in
detail in \cite{Buras:1991jm}. The second term gives the LH
contribution to the mass difference. As these are the corrections to
the SM contribution, we do not  
consider QCD corrections to them which would arise from gluonic loops added
to the diagrams of Fig \ref{fig:1}. The effective Lagrangian in the LH
model to order $\frac{v^2}{f^2}$\footnote{$f$ is the scale at which
the global SU(5) symmetry is spontaneously broken via a vacuum
expectation value which is expected to be in the TeV range and roughly
of the order of masses of heavy bosons and $v$ is the vev of standard
model Higgs} is well approximated by :
\beq
{\cal L}_{eff}^{\bigtriangleup J} = \frac{G_F^2}{16 \pi^2} M_W^2
S^{LH}_j Q(\bigtriangleup J = 2) 
\label{eq:3}
\eeq
where $J = B, S$ and $j=b, s$ for $B_d - \bar{B}_d$ and $K^0 -
\bar{K^0}$ respectively. They are given as :
\beqa
Q(\bigtriangleup B = 2) &=& (\bar{b}_\alpha d_\alpha)_{V-A} 
(\bar{b}_\beta d_\beta)_{V-A}                 \nonumber           \\
Q(\bigtriangleup S = 2) &=& (\bar{s}_\alpha d_\alpha)_{V-A} 
(\bar{s}_\beta d_\beta)_{V-A}
\label{eq:4}
\eeqa
and 
\beqa
S_b^{LH} &=& \frac{v^2}{f^2}
\Bigg[ 
\left\{ \sum_{i=u,c,t} \xi_i^2 E(x_i,W_L) 
+ \sum_{i\ne j = u,c,t,T} \xi_{ij} E(x_i,x_j,W_L) \right\} \nonumber \\
&& + \frac{2 c^2}{s^2} 
   \left\{ \sum_{i=u,c,t} {\xi'}_i^2 E(x_i,W_L,W_H)
     + \sum_{i\ne j=u,c,t,T} {\xi'}_{ij} E(x_i,x_j,W_L,W_H) \right\}
                        \nonumber \\
&& + \left( 1 - \frac{2 s_+ f}{v} \right) 
     \left\{ \sum_{i=u,c,t} \lambda_i^2 E(x_i,W_L,\Phi)
       + \sum_{i\ne j =u,c,t} \lambda_i \lambda_j 
               E(x_i,x_j,W_L,\Phi)
     \right\}
\Bigg]
\label{eq:5}
\eeqa
where $\xi_{i}, \xi_{ij}$ and functions $E$ are defined in
appendix \ref{appendix:a}. $x_i = m_i^2/m_W^2$, $\lambda$'s are the
CKM factors defined as  
$\lambda_i = V_{id} V^*_{ib} ~(i = u,c,t)$ and $\lambda_T =
\frac{\lambda_1}{\lambda_2} \lambda_t$ and $V_{ij}$'s are the CKM
matrix elements. 

%%%%%%%%%%%%%%%%%%%%%%
%%%%%%%%%%%%%%%%%%%%%%%

We note that despite the occurrence of spinless Higgs couplings to
quarks, the ultimate structure of the effective Hamiltonian in LH
retains the same $(V-A)$ form as in SM to order
$\left(\frac{v}{f}\right)^2$ . Given the form of the effective 
Hamiltonian, we can proceed exactly as in SM and calculate its matrix
element between $K_0 - {\bar K}_0$ or $B_0 - \bar{B}_0$ states using
the vacuum saturation approximation. There are no divergences in the
SM amplitude because of the unitarity of the CKM
matrix ; this statement holds even in LH model\footnote{the CKM matrix
is unitary in LH upto order $v^2/f^2$} where once again the 
unitarity of CKM ensures that all divergences vanish to order
$(v/f)^2$. Neglecting QCD corrections 
and long distance contributions we can get the mass difference to be :
\beq
\bigtriangleup M(B_0 - \bar{B_0})^{LH} = \frac{G_F^2}{6 \pi^2} M_B
M_W^2 f_B^2 S_b^{LH} 
\label{eq:6} 
\eeq
and 
\beq
\bigtriangleup M(K_0 - \bar{K_0})^{LH} = \frac{G_F^2}{6 \pi^2} M_K
M_W^2 f_K^2 S_s^{LH} 
\label{eq:7} 
\eeq
where $M_{B,K}, f_{B,K}$ are the masses and decay constants of B and K
mesons respectively \footnote{The QCD corrections to these have been
worked out in literature}.

\par It should be mentioned that the renormalization group 
evolution of the matrix elements
has been the subject of much work and has been summarized in
\cite{Bijnens:1990mz} and is far from trivial since the matrix elements
are controlled by long distance dynamics and are generally
parameterized by a {\sl ``Bag factor''} $B_{K,B}$. However for the
neutral B meson case, the long range interactions arising from the
intermediate virtual states are negligible because of the large B mass,
being far from the region of hadronic resonances.  

\par The LH involves not only heavy vector bosons and quarks but also
a large number of parameters over and above those in the SM. The
global symmetry in the theory is broken at TeV range scale
$\Lambda_s$($\Lambda_s = 4 \pi f$); the scalar bosons, doublets and
triplets, acquire  vacuum expectation values $v$ and $v'$ respectively
at the 
EW-scale, providing the convenient 
small parameters $v/f$ and $v'/f$. The mixing of the charged and
neutral vector bosons results in two mixing angle parameters $\theta$
and $\theta'$ (with $c = Cos\theta, s = Sin\theta, c' =
Cos\theta'$ and $s' = Sin\theta'$). Finally the Yukawa
coupling of the fermions involves two parameters $\lambda_1$ and
$\lambda_2$ with the combination $x_L =
\frac{\lambda_1^2}{\lambda_1^2 + \lambda_2^2}$ occurring
frequently. To the leading order in $(v/f)$, the masses of all the heavy
particles in LH can be expressed in terms of SM masses $m_W$ and $m_Z$
as :
\beqa
\frac{m_{W_H}}{m_W} &\approx& \frac{1}{s c} \frac{f}{v}. \nonumber \\
\frac{m_{T}}{m_t} &\approx &
\frac{\lambda_1^2 + \lambda_2^2}{\lambda_1 \lambda_2} \frac{f}{v}.
\nonumber \\ 
\frac{m_{\Phi}}{m_H} &\approx & \sqrt{2} \frac{f}{v}. \nonumber 
\eeqa

The coupling of all heavy particles to SM particles as well among
themselves are expressible in terms of these parameters with the SM
ones.

\par The parameter space is obviously too large. Requiring that the
heavy particles have masses in TeV range results in the condition $\frac{1}{s c}
< 10$. There is another restriction arising out of the requirement that
the mass of the triplet scalars be positive definite
\cite{Han:2003wu}:
\beq
\frac{v'^2}{v^2} < \frac{v^2}{16 f^2}
\nonumber 
\eeq
We have varied $v/f$ in the range 0 to 0.1. $s, 
s'$ in range 0.2 to 0.8 and $x_L$ in range 0.2 to 0.8 in our
numerical analysis. Other parameters used are given in the Appendix
\ref{appendix:b}.

%%%%%%%%%%%%%%%%%%%%%%%%%%%%%%%%%%%%%%%%%%%%%%%%%%%%%%%%%%%%%%
\begin{figure}[t]
\begin{center}
\epsfig{file=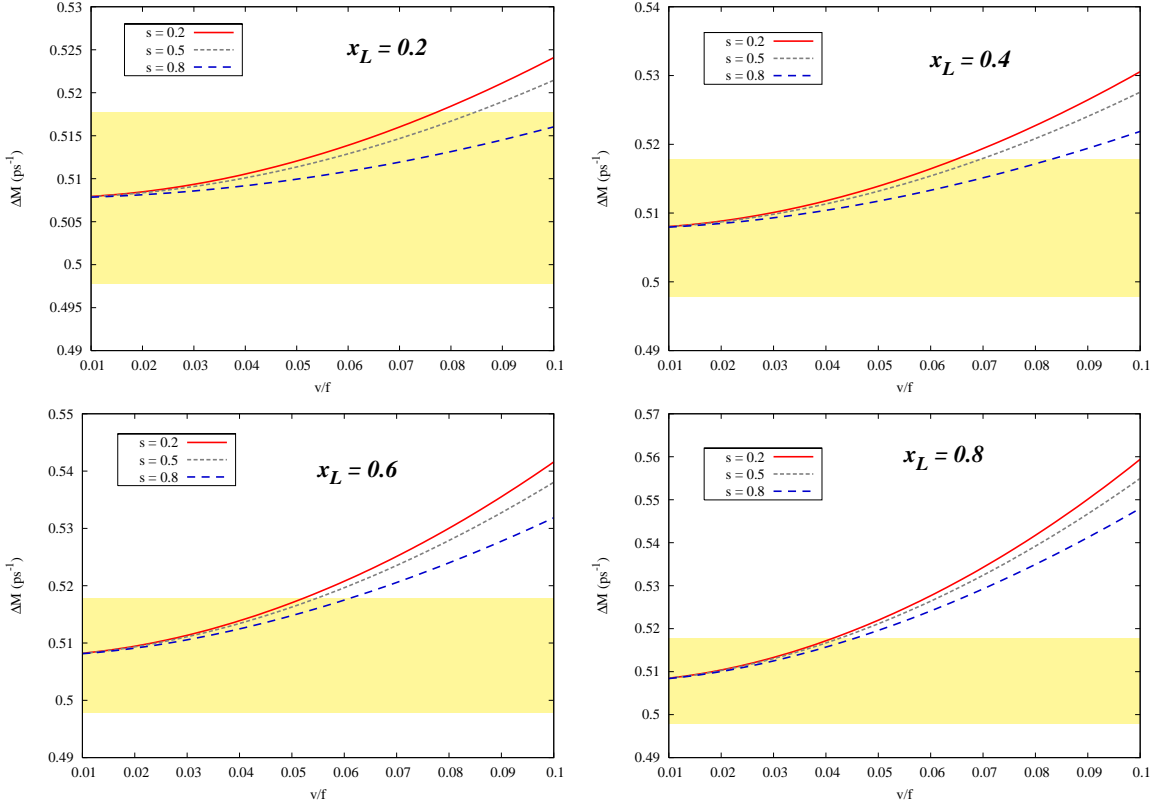,width=.9\textwidth}
\caption{$\bigtriangleup M(B_d - \bar{B}_d)$ in $ps^{-1}$ with
$v/f$. For these plots we have used $s' = s$. Shaded area indicates
the experimental bounds.}   
\label{fig:2}
\end{center}
\end{figure}
%%%%%%%%%%%%%%%%%%%%%%%%%%%%%%%%%%%%%%%%%%%%%%%%%%%%%%%%%%%%%%%%

\par Our results for the $B_d^0 - \bar{B_d^0}$ case are shown in
Figure \ref{fig:2}. Varying
$s'$ doesn't significantly change our conclusions. The
corresponding $K^0 - \bar{K^0}$ results too have similar
trend. However, since there are large error bars in them because of
QCD corrections involved, it makes it difficult to draw any definitive
conclusions. Hence we haven't shown them. In the plots shown in
Fig.\ref{fig:2} the shaded area corresponds to the mass difference
$\bigtriangleup M(B_d - \bar{B}_d) = 0.5 \pm 0.01 ~ {\rm ps^{-1}}$
which is consistent with current experimental bounds
\cite{Eidelman:2004wy}. 

\par From Fig \ref{fig:2}, it is easy to note that for low values of
$x_L$, there is a region of parameter space (in terms of the
parameters $s$ and $v/f$) that is consistent with the experimental
limits. Very specifically  for $x_L~=~0.2$, the bound on the scale $f$
can be very close to $1$  TeV for almost all the $s$ values. However,
as $x_L$ is increased, the LH contribution starts deviating
significantly from the SM results. Therefore, in principle the
experimentally allowed band for the $B_d^0 - \bar{B_d^0}$ mass
difference provides significant constraints for the parameter space of
generic Little Higgs models, and in particular the Littlest Higgs
model.  

\par It will be fruitful to compare the limits on the parameters
coming from precision electroweak data \cite{precision}. To this end
we recall that the Littlest Higgs model does not have the {\it
custodial symmetry} inherently built into it and can therefore, in
principle, lead to large corrections, arising both from heavy gauge
boson exchange diagrams and the triplet VEV. A naive way out would be
to have the extra gauge boson masses raised by some means. However,
this would spoil the motivation of circumventing the hierarchy problem
and would also bring in the issue of fine tuning. It has been found
that global fits to precision electroweak data imply the following
bound (at $95\%$ C.L.) on the scale $f$ for any generic coupling
(specifically varying $c$ and $c'$ between $0.1$ and $0.995$): 
\beq
f > 4 ~TeV
\eeq
It is worth noting that this very stringent bound is almost
(practically) independent of any variation of Higgs mass upto $200$
GeV. Further, the bound still holds for any order unity value of the 
parameters encapsulating the physics due to proper UV-completion of
the theory. To be noted is the fact that these are very strong limits
and the origin can simply be traced back to the absence of custodial
symmetry. In the second reference of \cite{precision} it was noted
that considering only precision electroweak data allows for a small
region in parameter space where the bound on the scale $f$ can be
lowered to about $1$ TeV. However, electroweak data combined with
Drell-Yan production excludes this region and a combined analysis
implies a bound very similar to the one quoted above. Constraints from
low energy precision data like $(g-2)_{\mu}$ and atomic weak charge of
Cesium also indicate similar bounds, though it should be remembered 
that the rather small $(g-2)_{\mu}$ corrections may not serve to put
any meaningful bounds.

\par Interestingly enough, in almost all the variations of the
Littlest Higgs model \cite{lhvariation}, the constraints remain quite
strong and generically very similar to the minimal version, though for
some very specific choices of the parameters the constraints on $f$
are relaxed to $1-2$ TeV. This can be understood as arising due to
small mixing between the two sets of gauge bosons and also small
coupling between the fermions and heavy $U(1)$ gauge boson. 
Nevertheless, these arise only in very specific models and for very
special choices of the parameters and are not a generic feature of LH
models. It may also be useful to keep in mind that the positivity of
triplet mass squared imposes severe constraints on the triplet VEV and
therefore the parameters or combination of parameters entering the
mass squared relation. This strong constraint considerably reduces the
allowed parameter space. Therefore, it is natural to expect that in
the variation of the minimal model where there is no triplet Higgs,
the bounds are partially relaxed as is the case in models which have
custodial symmetry built into them. 

\par Turning to contribution to $\bigtriangleup M (B_d - \bar{B}_d)$,
we would indeed get a bound similar to eqn (9) above if we require
that the LH contribution be no more than the experimental band for  
$\bigtriangleup M (B_d - \bar{B}_d)$. 
However, as has been pointed out by Buras {\sl et.al.}
\cite{Buras:2004kq} that there is  a hadronic uncertainity of about
10\% in calculation of $\bigtriangleup M (B_d - \bar{B}_d)$. In view
of this a reasonable constraint on $(v/f)$ could be given if the
variation in the mass difference could be more than 10\%. For a
contribution of about 10\% , the LH model would require $v/f \le 0.2$,
which is not very useful in view of the stronger constraint like eqn(9)
above. However, should it become possible for hadronic uncertainities
to be reduced by a factor of 2 or 3, then the bound on $(v/f)$ values
becomes much lower, leading to constraints on the value of $f$
comparable with the value in equation (9) above.

%%%%%%%%%%%%%%%%%%%%%%%%%%%
%  Acknowledgements

\section*{Acknowledgments}
We thank Heather Logan for useful discussions. We thank 
Andrzej Buras for  useful communications regarding their paper. 
We thank him and Selma Uhlig for discussions regarding this revised 
version of our paper. The work of SRC, AG
and NG is supported under the SERC scheme of Department of Science \&
Technology (DST), India under the project no. SP/S2/K-20/99.   

%%%%%%%%%%%%%%%%%%%%%%%%%%%
%  Appendix 

\appendix

\renewcommand{\theequation}{\thesection.\arabic{equation}}
\setcounter{equation}{0}

\section{Loop Functions \label{appendix:a}}
$W_L, W_H$ in eqn (\ref{eq:5}) refer to the light \& heavy W-boson in LH;
c,s are the mixing angles in LH. Various $\xi$'s are 
\beqa
\xi_i^2 &=& 2 c^2 (c^2 - s^2) \lambda_i^2  , ~~ i = u,c
\label{app:a:1} \\
\xi_t^2 &=& 2 \left\{c^2 (c^2 - s^2) + x_L^2 \right\} \lambda_t^2 
\label{app:a:2} \\ 
\xi_{it} &=& 2 \left\{ c^2 (c^2 - s^2) + \frac{x_L^2}{2} \right\} 
\lambda_i \lambda_t   \label{app:a:3} \\
\xi_{iT} &=& - x_L^2 \lambda_i \lambda_t       \label{app:a:4} \\
\xi_{ij} &=& \xi_{ji} = \xi_i \xi_j   \label{app:a:4(a)}  \\
\xi_{tT} &=& - x_L^2 \lambda_t^2        \label{app:a:4(b)}  \\
{\xi'}_i^2 &=& - (f/v)^2 \lambda_i^2  \label{app:a:4(c)}  \\
{\xi'}_{ij} &=& - (f/v)^2 \lambda_{ij}  \label{app:a:4(d)} 
\eeqa
the functions ($E$) used in eqn. (\ref{eq:5}) are :
\beqa
E(x_i,x_j,W_L) &=& - \frac{x_i x_j}{x_i - x_j} 
\left\{  {1 \over 4} + \frac{3}{2} \frac{1}{(1 - x_i)}
   - \frac{3}{4} \frac{1}{(1 - x_i)^2} 
\right\} log(x_i)                                 \nonumber \\
&& - \frac{x_j x_i}{x_j - x_i} 
\left\{  {1 \over 4} + \frac{3}{2} \frac{1}{(1 - x_j)}
   - \frac{3}{4} \frac{1}{(1 - x_j)^2} 
\right\} log(x_j)                                 \nonumber \\ 
&&+ \frac{3}{4} \frac{x_i x_j}{(1 - x_i)(1 - x_j)}
\label{app:a:5}  \\
E(x_i,W_L) &=&  - \frac{3}{2} \left(\frac{x_i}{x_i-1}\right)^3
log(x_i) 
- x_i \left\{ {1\over 4} + \frac{9}{4} \frac{1}{(1 - x_i)}
- {3 \over 2} \frac{1}{(1 - x_i)^2} 
\right\}  \label{app:a:6}  \\
E(x_i, x_j, W_L, W_H) &=&  - \frac{x_i x_j}{x_{W_H} (x_i - x_j) (1 -
x_i) (1 - \frac{x_i}{x_{W_H}})} \left\{ 1 - 
\left(1 + \frac{1}{x_{W_H}}\right) x_i + \frac{x_i^2}{4 x_{W_H}}
\right\} log(x_i)                                 \nonumber \\ 
&& - \frac{x_j x_i}{x_{W_H} (x_j - x_i) (1 -
x_j) (1 - \frac{x_j}{x_{W_H}})} \left\{ 1 - 
\left(1 + \frac{1}{x_{W_H}}\right) x_j + \frac{x_j^2}{4 x_{W_H}}
\right\} log(x_j)   \nonumber \\
&& + {3 \over 4} \frac{1}{\left(1 - \frac{1}{x_{W_H}}\right) 
\left(1 - \frac{x_i}{x_{W_H}}\right) \left(1 -
\frac{x_j}{x_{W_H}}\right)} log(x_{W_H})   
\label{app:a:7}   \\
E(x_i, W_L, W_H) &=& {3 \over 4} \frac{x_i^3}{x_{W_H}^2 (1 - x_i)^2
\left(1 - \frac{x_i}{x_{W_H}}\right)^2} 
\left\{2 - \frac{1 + x_{W_H}}{x_{W_H}} x_i \right\} log{x_i} 
                              \nonumber \\
&& + {3 \over 4} \frac{1}{\left(1 - \frac{1}{x_{W_H}}\right)
\left(1 - \frac{x_i}{x_{W_H}}\right)^2} log(x_{W_H}) \nonumber \\
&& - \frac{x_i}{x_{W_H} (1 - x_i) \left(1 -
\frac{x_i}{x_{W_H}}\right)} \left\{1 - \left(1 +
\frac{1}{x_{W_H}}\right) x_i + \frac{x_i^2}{4 x_{W_H}} \right\}
\label{app:a:8}    \\
E(x_i,x_j,x_{W_L},x_\phi) &=& \frac{x_i x_j}{2}
\left\{ - \frac{x_i \left(1 - \frac{x_i}{4}\right)}{(x_i - x_j) (1 -
x_i) (x_\phi - x_i)} log(x_i)  
 - \frac{x_j \left(1 - \frac{x_j}{4}\right)}{(x_j - x_i) (1
- x_j) (x_\phi - x_j)} log(x_j)   \right. \nonumber \\  
&& \left. + \frac{x_\phi \left(1 - \frac{x_\phi}{4}\right)}{(x_\phi -
x_i) (x_\phi - x_j) (1 - x_\phi)} log(x_\phi) \right\}    
\label{app:a:9}    \\
E(x_i, x_{W_L}, x_\phi) &=&  \frac{x_i^2}{2}
\Bigg[ - \frac{\left(1 - \frac{x_i}{4}\right)}{(1 - x_i)(x_\phi - x_i)} 
- \frac{\left\{ x_\phi \left(1 - \frac{x_i}{2}\right) + \frac{3
x_i^2}{4} \left(\frac{x_\phi}{3} - 1\right) \right\}}{(1 - x_i)^2
(x_\phi - x_i)^2} log(x_i)  \nonumber \\
&& + \frac{x_\phi \left(1 - \frac{x_\phi}{4}\right)}{(x_\phi - x_i)^2
(1 - x_\phi)} log(x_\phi) \Bigg]
\eeqa

\section{\label{appendix:b}Input parameters}

$$G_F = 1.16 \times 10^{-5} {\rm GeV}^{-2} ~,~ f_B = 0.21
~,~ m_B = 5.3 {\rm GeV}, $$
$$m_{W_L} = 80.4 {\rm GeV} ~,~ m_{Z_L} = 91.2 {\rm GeV} $$

%%%%%%%%%%%%%%%%%%%%%%%%%%%%%%%%%
%  Refrences 

%%%%%%%%%%%%%%%%%%%%%%%%%%%


\begin{thebibliography}{99}

%\cite{Buras:2004kq}
\bibitem{Buras:2004kq}
A.~J.~Buras, A.~Poschenrieder and S.~Uhlig,
%``Particle antiparticle mixing, epsilon(K) and the unitarity triangle in the
%littlest Higgs model,''
arXiv:hep-ph/0410309.
%%CITATION = HEP-PH 0410309;%%


%%%%%%%%%%%%%%%%%%%%%%%%%%%

%\cite{Arkani-Hamed:2001nc}
\bibitem{Arkani-Hamed:2001nc}
 N.~Arkani-Hamed, A.~G.~Cohen and H.~Georgi,
 %``Electroweak symmetry breaking from dimensional deconstruction,''
 Phys.\ Lett.\ B {\bf 513}, 232 (2001)
 [arXiv:hep-ph/0105239]; 
 %%CITATION = HEP-PH 0105239;%%
%
%\cite{Arkani-Hamed:2002pa}
%\bibitem{Arkani-Hamed:2002pa}
 N.~Arkani-Hamed, A.~G.~Cohen, T.~Gregoire and J.~G.~Wacker,
 %``Phenomenology of electroweak symmetry breaking from theory space,''
 JHEP {\bf 0208}, 020 (2002)
 [arXiv:hep-ph/0202089] ; 
 %%CITATION = HEP-PH 0202089;%%
%
%\cite{Arkani-Hamed:2002qx}
%\bibitem{Arkani-Hamed:2002qx}
 N.~Arkani-Hamed, A.~G.~Cohen, E.~Katz, A.~E.~Nelson, T.~Gregoire and J.~G.~Wacker,
 %``The minimal moose for a little Higgs,''
 JHEP {\bf 0208}, 021 (2002)
 [arXiv:hep-ph/0206020] ;
 %%CITATION = HEP-PH 0206020;%%
%
%\cite{Arkani-Hamed:2002qy}
%\bibitem{Arkani-Hamed:2002qy}
 N.~Arkani-Hamed, A.~G.~Cohen, E.~Katz and A.~E.~Nelson,
 %``The littlest Higgs,''
 JHEP {\bf 0207}, 034 (2002)
 [arXiv:hep-ph/0206021] ; 
 %%CITATION = HEP-PH 0206021;%%
%
%\cite{Low:2002ws}
%\bibitem{Low:2002ws}
 I.~Low, W.~Skiba and D.~Smith,
 %``Little Higgses from an antisymmetric condensate,''
 Phys.\ Rev.\ D {\bf 66}, 072001 (2002)
 [arXiv:hep-ph/0207243].
 %%CITATION = HEP-PH 0207243;%%

%%%%%%%%%%%%%%%%%%%%%%%%%%%

%\cite{Han:2003wu}
 \bibitem{Han:2003wu}
 T.~Han, H.~E.~Logan, B.~McElrath and L.~T.~Wang,
 %``Phenomenology of the little Higgs model. ((U)),''
 Phys.\ Rev.\ D {\bf 67}, 095004 (2003)
 [arXiv:hep-ph/0301040].
 %%CITATION = HEP-PH 0301040;%%

%%%%%%%%%%%%%%%%%

%\cite{Schmaltz:2002wx}
\bibitem{Schmaltz:2002wx}
 M.~Schmaltz,
 %``Physics beyond the standard model (Theory): Introducing the little Higgs,''
 Nucl.\ Phys.\ Proc.\ Suppl.\  {\bf 117}, 40 (2003)
 [arXiv:hep-ph/0210415] ; 
 %%CITATION = HEP-PH 0210415;%%
%
%\cite{Logan:2003sf}
%\bibitem{Logan:2003sf}
 H.~E.~Logan,
 %``Phenomenology of the littlest Higgs model,''
 arXiv:hep-ph/0307340 ; 
 %%CITATION = HEP-PH 0307340;%%
%
%\cite{Logan:2003pa}
%\bibitem{Logan:2003pa}
 H.~E.~Logan,
 %``Little Higgs phenomenology,''
 arXiv:hep-ph/0310151.
 %%CITATION = HEP-PH 0310151;%%

%%%%%%%%%%%

%\cite{Chen:2003fm}
\bibitem{Chen:2003fm}
 M.~C.~Chen and S.~Dawson,
 %``One-loop radiative corrections to the rho parameter in the littlest Higgs
%model,''
 arXiv:hep-ph/0311032 ;
 %%CITATION = HEP-PH 0311032;%%
%
%\cite{Yue:2004xt}
%\bibitem{Yue:2004xt}
 C.~x.~Yue and W.~Wang,
 %``The branching ratio R(b) in the littlest Higgs model,''
 Nucl.\ Phys.\ B {\bf 683}, 48 (2004)
 [arXiv:hep-ph/0401214].
 %%CITATION = HEP-PH 0401214;%%
%
%\cite{Kilian:2003xt}
%\bibitem{Kilian:2003xt}
 W.~Kilian and J.~Reuter,
 %``The low-energy structure of little Higgs models,''
 arXiv:hep-ph/0311095.
 %%CITATION = HEP-PH 0311095;%%

%%%%%%%%%%%

%\cite{Huo:2003vd}
\bibitem{Huo:2003vd}
 W.~j.~Huo and S.~h.~Zhu,
 %``b $\to$ s gamma in littlest Higgs model,''
 Phys.\ Rev.\ D {\bf 68}, 097301 (2003)
 [arXiv:hep-ph/0306029].
 %%CITATION = HEP-PH 0306029;%%

%%%%%%%%%%%%%

%\cite{Han:2003gf}
\bibitem{Han:2003gf}
 T.~Han, H.~E.~Logan, B.~McElrath and L.~T.~Wang,
 %``Loop induced decays of the little Higgs: H $\to$ g g, gamma gamma,''
 Phys.\ Lett.\ B {\bf 563}, 191 (2003)
 [arXiv:hep-ph/0302188].
 %%CITATION = HEP-PH 0302188;%%

%%%%%%%%%%%%%%

%\cite{Choudhury:2004ce}
\bibitem{Choudhury:2004ce}
 S.~R.~Choudhury, N.~Gaur, G.~C.~Joshi and B.~H.~J.~McKellar,
 %``K(L) $\to$ pi0 nu anti-nu in little Higgs model,''
 arXiv:hep-ph/0408125.
 %%CITATION = HEP-PH 0408125;%%

%%%%%%%%%%%%%%%%%%%%%%%%%

%\cite{Buras:1991jm}
\bibitem{Buras:1991jm}
 A.~J.~Buras, M.~Jamin, M.~E.~Lautenbacher and P.~H.~Weisz,
 %``Effective Hamiltonians for Delta S = 1 and Delta B = 1 nonleptonic decays
 %beyond the leading logarithmic approximation,''
 Nucl.\ Phys.\ B {\bf 370}, 69 (1992)
 [Addendum-ibid.\ B {\bf 375}, 501 (1992)];
 %%CITATION = NUPHA,B370,69;%%
%
%%\cite{Buras:1993dy}
%\bibitem{Buras:1993dy}
A.~J.~Buras, M.~Jamin and M.~E.~Lautenbacher,
%``The Anatomy of epsilon-prime / epsilon beyond leading logarithms with
%improved hadronic matrix elements,''
Nucl.\ Phys.\ B {\bf 408}, 209 (1993)
[arXiv:hep-ph/9303284] ; 
%%CITATION = HEP-PH 9303284;%%
%
%%\cite{Bigi:yz}
%\bibitem{Bigi:yz}
I.~I.~Y.~Bigi and A.~I.~Sanda,
%``CP Violation,''
Cambridge Monogr.\ Part.\ Phys.\ Nucl.\ Phys.\ Cosmol.\  {\bf 9}, 1
(2000). 
%%CITATION = CMPCE,9,1;%%

%%%%%%%%%%%%%%%%%%%%%%%%%%

%\cite{Bijnens:1990mz}
\bibitem{Bijnens:1990mz}
 J.~Bijnens, J.~M.~Gerard and G.~Klein,
 %``The K(L) - K(S) Mass Difference,''
 Phys.\ Lett.\ B {\bf 257}, 191 (1991) ; 
 %%CITATION = PHLTA,B257,191;%%
%
%\cite{Buras:1996cw}
%\bibitem{Buras:1996cw}
 A.~J.~Buras,
 %``Theoretical review of K physics,''
 arXiv:hep-ph/9609324.
 %%CITATION = HEP-PH 9609324;%%


%%%%%%%%%%%%%

%\cite{Eidelman:2004wy}
\bibitem{Eidelman:2004wy}
 S.~Eidelman {\it et al.}  [Particle Data Group Collaboration],
 %``Review of particle physics,''
 Phys.\ Lett.\ B {\bf 592}, 1 (2004).
 %%CITATION = PHLTA,B592,1;%%

%%%%%%%%%%%%%%%%%

%\cite{precision}
\bibitem{precision}
C.~Csaki, J.~Hubisz, G.~D.~Kribs, P.~Meade and J.~Terning,
%``Big corrections..''
Phys.\ Rev.\ D {\bf 67}, 115002 (2003);
 %%CITATION = PHRVA,D67,115002;%%
%
J.~L.~Hewett, F.~J.~Petriello and T.~G.~Rizzo,
%``Constraining the Littlest Higgs''
JHEP {\bf 0310}, 062 (2003).
%%CITATION = HEP-PH 0211218;%%

%%%%%%%%%%%%%%%%%%%%%%%

%\cite{precision}
\bibitem{lhvariation}
C.~Csaki, J.~Hubisz, G.~D.~Kribs, P.~Meade and J.~Terning,
%``Variations of Little Higgs ....''
Phys.\ Rev.\ D {\bf 68}, 035009 (2003);
 %%CITATION = PHRVA,D68,035009;%%

\end{thebibliography}
\end{document}